\documentclass[prd,nofootinbib,notitlepage,superscriptaddress]{revtex4-1}
\usepackage{epsfig,amsmath,slashed}

\newcommand{\di}{{\rm d}}
\newcommand{\ii}{i}
\newcommand{\Pro}{{\sf P}}
\def\spt{{\cal S}}
\def\wT{{\widehat T}}
\def\wspt{{\widehat{\cal S}}}

\def\wpsi{{\widehat{\psi}}}
\def\wrho{{\widehat{\rho}}}

\def\codevmu{{\stackrel{\leftrightarrow}{\partial^\mu}}}
\def\codevnu{{\stackrel{\leftrightarrow}{\partial^\nu}}}
\newcommand{\tr}{{\rm tr}}  
  
\newcommand{\e}{{\rm e}}

\newcommand{\omegav}{\boldsymbol{\omega}}
\newcommand{\betav}{\boldsymbol{\beta}}

\newcommand{\sigmav}{\boldsymbol{\sigma}}

\newcommand{\Piv}{\boldsymbol{\Pi}}
\newcommand{\p}{{\rm p}}
\newcommand{\x}{{\rm x}}

\newcommand{\Psibar}{{\overline \Psi}}

\newcommand{\omt}{\frac{\omega}{T}}
\newcommand{\chiomt}{\chi\!\left( \omt \right)}
\newcommand{\be}{\begin{equation}}
\newcommand{\ee}{\end{equation}}                                                                               
\newcommand{\bea}{\begin{eqnarray}}
\newcommand{\eea}{\end{eqnarray}}                                                                               

\begin{document}

\begin{center}
\Large\bf{Relativistic distribution function for particles with spin
at local thermodynamical equilibrium}
\end{center} 
\medskip
\begin{center} 
F. Becattini \\
\vspace{0.3 cm}
{\it Universit\`a di Firenze and INFN Sezione di Firenze, Florence, Italy \\
Universit\"at Frankfurt and FIAS, Frankfurt am Main, Germany} \\
e-mail: becattini@fi.infn.it \\
\vspace{0.3 cm}
V. Chandra \\
\vspace{0.3 cm}
{\it INFN Sezione di Firenze, Florence, Italy} \\
e-mail: vinod.chandra@fi.infn.it \\
\vspace{0.3 cm}
L. Del Zanna, E. Grossi \\
\vspace{0.3 cm}
{\it Universit\`a di Firenze and INFN Sezione di Firenze, Florence, Italy} \\
e-mail: ldz@arcetri.astro.it and grossi@fi.infn.it
\end{center}

\begin{abstract}
We present an extension of relativistic single-particle distribution function for 
weakly interacting particles at local thermodynamical equilibrium including spin 
degrees of freedom, for massive spin 1/2 particles. We infer, on the basis of the
global equilibrium case, that at local thermodynamical equilibrium particles acquire 
a net polarization proportional to the vorticity of the inverse temperature four-vector 
field. The obtained formula for polarization also implies that a steady gradient of 
temperature entails a polarization orthogonal to particle momentum. The single-particle
distribution function in momentum space extends the so-called Cooper-Frye formula to particles
with spin 1/2 and allows to predict their polarization in relativistic heavy ion 
collisions at the freeze-out. 
\end{abstract} 

\maketitle

\section{Introduction}
\label{intro}

The single-particle distribution function is the main quantity in kinetic theory 
and its form at local thermodynamical equilibrium for relativistic, weakly interacting,
gases is well known. For spinless particles, it is simply the Bose-Einstein distribution 
with $x$-dependent values of temperature and chemical potential (which can be defined
as Bose-J\"uttner distribution):
\be\label{localf}
  f(x,p) = \frac{1}{\e^{\beta(x) \cdot p - \xi(x) } - 1}
\ee
where $\beta = \frac{1}{T_0} u$ is the inverse temperature four-vector,$T_0$ being 
the proper temperature measured by a comoving thermometer with the four-velocity $u$ and 
$\xi=\mu_0/T_0$ is the ratio between the proper chemical potential $\mu_0$ and $T_0$. 
The above formula has, as straightforward consequence, the invariant momentum spectrum 
at local thermodynamical equilibrium, the so-called Cooper-Frye formula \cite{cf}:
\be\label{cooper}
 \varepsilon \frac{\di N}{\di^3 \p} = \int_\Sigma \di \Sigma_\mu p^\mu f(x,p)
\ee
where $\Sigma$ is a space-like 3-dimensional hypersurface. This formula is widespreadly
used in e.g. relativistic heavy ion collisions to calculate hadronic spectra at the end
of the hydrodynamical stage.

The distribution (\ref{localf}), multiplied by a degeneracy factor $(2S+1)$, 
is also used for particles with spin (for fermions replacing the $-1$ with $+1$ in 
the denominator) being understood that $f$ means the {\em total} particle density 
in phase space, i.e. summed over polarization states. However, in general, particles 
may not evenly populate the various polarization states and one may then wonder what is 
the appropriate extension of (\ref{localf}) in this case. Indeed, in this work, we 
will answer this question and provide a generalization of (\ref{localf}) and (\ref{cooper})
including the spin degrees of freedom. We will argue that 
a non-even population of the polarization states arises when the inverse temperature 
four-vector field has a non-vanishing antisymmetric part of its gradient and calculate the 
polarization vector for massive spin $1/2$ particles. Phenomenologically, this extension may
have several interesting applications. For instance, it would make it possible to 
predict the value of particle polarization in relativistic heavy ion collisions 
\cite{sorin,wang,wang2,torrieri,csernai} at the hydrodynamical decoupling, provided that 
local thermodynamical equilibrium applies to spin degrees of freedom as well. 

The derivation of the extension of Cooper-Frye formula will be done in several steps and
it also requires a summarization of relativistic kinetic theory from quantum field 
viewpoint including spinorial degrees of freedom. The reader who is only interested 
in the final result may jump to sect.~\ref{conclu}.

The paper is organized as follows: in sect.~\ref{kinetics} we set the stage for the
generalization of the single-particle distribution function according to relativistic
kinetic theory in a quantum field framework; in sect.~\ref{global} we summarize results 
obtained for global thermodynamical equilibrium with rotation deriving more compact formulae
for spin $1/2$ particles; in sect.~\ref{local} we generalize the single-particle
distribution function for spin $1/2$ particles to local thermodynamical equilibrium
and discuss the physical meaning of the antisymmetric tensor coupled to the spin
matrices; in sect.~\ref{polarization} we calculate the polarization vector in a 
relativistic fluid and in sect.~\ref{conclu} we summarize the results and draw 
conclusions.

\subsection*{Notation}

In this paper we adopt the natural units, with $\hbar=c=K=1$.\\
The Minkowskian metric tensor is ${\rm diag}(1,-1,-1,-1)$; for the Levi-Civita
symbol we use the convention $\epsilon^{0123}=1$.\\ 
We will use the relativistic notation with repeated indices assumed to 
be saturated. Operators in Hilbert space will be denoted by an upper hat, e.g. 
$\widehat {\sf R}$, with the exception of the Dirac field operator which is 
denoted with a capital $\Psi$. Repeated greek indices are saturated, repeated
latin indices are not.

\section{Relativistic kinetic theory and quantum fields}
\label{kinetics}

Classically, hydrodynamics emerges as an effective description of the underlying 
microscopic dynamics, specifically of an underlying kinetic theory, where the 
substratum is a large number of interacting particles.
In this approach, all hydrodynamical quantities and, chiefly, the stress-energy 
tensor in relativistic hydrodynamics, can be expressed in terms of the single particle 
distribution function $f(x,p)$:
\be\label{tens1}
    T^{\mu \nu}(x) = \int \frac{\di^3 \p}{\varepsilon} \; p^\mu p^\nu f(x,p)
\ee    
where $f(x,p)$ fulfills a transport equation, notably Boltzmann equation.  
On the other hand, (relativistic) hydrodynamics can also be seen as the realization,
on the average, of conservation equations holding at the more fundamental level of 
quantum field operators. In this approach, the stress-energy tensor is seen as the 
mean value of the corresponding quantum one, which is a function of the fundamental fields:
\be\label{tens2}
  T^{\mu\nu}(x) = \tr (\wrho :\wT^{\mu \nu}(x):)
\ee  
being $\wrho$ is the density operator (either pure or mixed) and the normal ordering
is meant to remove unwanted divergencies \cite{degroot}; in general, a renormalization 
procedure is in order when calculating mean values in an interacting field theory 
\cite{callan}. 

What is the relation between the definitions (\ref{tens1}) and (\ref{tens2})? The 
connection between relativistic kinetic and quantum field theories has been nicely 
described in ref.~\cite{degroot} and the conceptual tool bridging the gap between them
is the {\em covariant Wigner function}. For a charged scalar field $\wpsi$ it is 
defined as:
\be\label{scawig}
 W(x,k) = \frac{1}{(2\pi)^4} \int \di^4 y \; 2 \, \e^{-\ii k \cdot y} 
 \langle : \wpsi^\dagger(x+y/2) \, \wpsi(x-y/2): \rangle
\ee                                                                   
where $\langle \, . \, \rangle$ stands for $\tr (\wrho \, . )$.  
Provided that the interaction terms are small and that significant $W(x,p)$ variations
occur on a macroscopic scale, i.e. much larger than Compton wavelength, $k$ is an 
almost on-mass-shell four vector and the covariant Wigner function can be written:
\bea\label{wigvsf}
 && W^+(x,k) \equiv \theta(k^0) W(x,k) = \int \frac{\di^3 \p}{\varepsilon} 
  \; \delta^4(k-p) f(x,p)
  \nonumber \\
 && W^-(x,k) \equiv \theta(-k^0) W(x,k) = \int \frac{\di^3 \p}{\varepsilon} \; 
  \delta^4(k+p) \bar f(x,p) 
\eea
which {\em define} the distribution functions $f(x,p)$ and $\bar f(x,p)$ with on-shell 
four-vector $p$ and $\varepsilon=\sqrt{{\bf p}^2+m^2}$. For a free field, the definition 
(\ref{scawig}) and the (\ref{wigvsf}) lead to:
\begin{eqnarray*}
 && f(x,p) = \frac{1}{2(2\pi)^3} \int \di^4 u \; \delta(u \cdot p) \, \e^{-\ii u \cdot x} 
 \langle a^\dagger_{p-u/2} a_{p+u/2} \rangle  
 \nonumber \\
 && \bar f(x,p) = \frac{1}{2(2\pi)^3} \int \di^4 u \; \delta(u \cdot p) \, 
 \e^{-\ii u \cdot x} \langle b^\dagger_{p-u/2} b_{p+u/2} \rangle  
\end{eqnarray*}
$a_p,b_p$ being destruction operators of particles with four-momentum $p$ normalized
so as to:
$$
  [a_p,a^\dagger_{p'}] = 2 \, \varepsilon \, \delta^3({\bf p}-{\bf p}')
$$
Thus:
\be\label{spectra}
 \int \di^3 \x \; f(x,p) = \frac{1}{2\varepsilon} \langle a^\dagger_{p} a_{p} \rangle  
 = \frac{\di N}{\di^3 p} \qquad \qquad 
 \int \di^3 x \; \bar f(x,p) = \frac{1}{2\varepsilon} \langle b^\dagger_{p} b_{p} \rangle 
 = \frac{\di \bar N}{\di^3 p} 
\ee
which is exactly what one would like to have for the non-interacting case, namely
the space integral of the distribution functions is the number of (anti)particles 
per three-momentum cell. 

The function $f$ turns out to be the familiar single-particle distribution function 
in phase space and, in the weakly interacting case, it can be shown to obey the 
Boltzmann equation \cite{degroot}. 

The covariant Wigner function makes the connection between (\ref{tens1}) and (\ref{tens2})
manifest. If the quantum stress energy tensor of the free scalar field is chosen to be:
$$
    \wT^{\mu\nu}(x) = -\frac{\ii}{2} \wpsi^\dagger(x) \codevmu \codevnu \wpsi(x)
$$
then \cite{degroot}:
\be\label{setvsf}
  T^{\mu\nu}(x) = \langle :\wT^{\mu \nu}(x): \rangle = -\frac{\ii}{2} \langle :
  \wpsi^\dagger(x) \codevmu \codevnu \wpsi(x): \rangle
  = \int \di^4 k \; k^\mu k^\nu W(x,k) = \int \frac{\di^3 \p}{\varepsilon} \; 
  p^\mu p^\nu \left[ f(x,p) + \bar f(x,p) \right]
\ee
where the eq.~(\ref{wigvsf}) has been used. We note in passing that a different 
form of the stress-energy tensor (differing thereof by a divergence) of the charged scalar 
field would have not led to the last, somewhat familiar, expression. We do not know 
whether a suitable change in the definition of $W$, $f$ or both, allows to keep relations 
(\ref{spectra}) and (\ref{setvsf}) for any form of the quantum stress-energy tensor. 
Also, it is worth pointing out that, in general, it has been shown that different 
quantum stress-energy tensors are thermodynamically inequivalent~\cite{bt1,bt2} and
it would not be thus surprising if a change of the quantum stress-energy tensor would 
not preserve the relation (\ref{setvsf}).

A similar connection can be built up for the spin 1/2 particles and the Dirac field.
In this case, the covariant Wigner function is a $4\times4$ spinorial matrix:
\bea\label{wigdir}
   W(x,k)_{AB} &=& - \frac{1}{(2\pi)^4} \int \di^4 y \; \e^{-\ii k \cdot y}
   \langle : \Psi_A (x-y/2) \Psibar_B (x+y/2) : \rangle \nonumber \\
   &=& \frac{1}{(2\pi)^4} \int \di^4 y \; \e^{-\ii k \cdot y} \langle : 
   \Psibar_B (x+y/2) \Psi_A (x-y/2): \rangle  
\eea
$\Psi$ being the Dirac field \footnote{It should be reminded that the normal ordering 
for fermion fields involves a minus sign for each permutation, e.g. $:a a^\dagger: 
= - a^\dagger a$. Therefore, taking into account anticommutation relations, for fields 
$:\Psi_A(x)\Psibar_B(y): = -:\Psibar_B(y) \Psi_A(x):$}.
We note in passing that the above definition has to be modified in full spinor 
electrodynamics \cite{elze,wang2} to preserve gauge invariance, but this can be
neglected for the scope of this work.
  
Likewise, the distribution functions, for on-shell $k$, are defined by an 
equation similar to (\ref{wigvsf}): 
\bea
 && W^+(x,k) \equiv \theta(k^0) W(x,k) = \frac{1}{2} 
 \int \frac{\di^3 \p}{\varepsilon} \; \delta^4(k-p) \sum_{r,s} u_{r}(p) f_{rs}(x,p)
  \bar u_{s}(p) \label{wigvsfd} \\
 && W^-(x,k) \equiv \theta(-k^0) W(x,k) = - \frac{1}{2} 
 \int \frac{\di^3 \p}{\varepsilon} \; \delta^4(k+p) \sum_{r,s} v_{s}(p) 
  \bar f_{rs}(x,p) \bar v_{r}(p) \label{wigvsfd.2}
\eea 
where $u_r(p),v_r(p)$ are the spinors of free particles and antiparticles in the 
polarization state $r$ (namely a helicity or third spin component) normalized
so as to $\bar u_r u_r = 2m$, $\bar v_r v_r = -2m$. Note that the distribution 
functions are $2 \times 2$ matrices, whose diagonal components $f_{rr}$ are the 
phase-space densities of particles in the polarization state $r$, that is (for the 
proof, see Appendix A):
\be\label{spectradirac}
 \int \di^3 \x \; f_{rr}(x,p) = \frac{1}{2\varepsilon} \langle a^\dagger_{p,r} 
 a_{p,r} \rangle  = \frac{\di N_r}{\di^3 p} \qquad \qquad 
 \int \di^3 x \; \bar f_{rr}(x,p) = \frac{1}{2\varepsilon} \langle b^\dagger_{p,r} 
 b_{p,r} \rangle = \frac{\di \bar N_r}{\di^3 p} 
\ee
It is now convenient to introduce a more compact notation, where $f$ is a $2 \times 2$
matrix and $U(p) = (u_+(p),u_-(p))$ a $4 \times 2$ matrix (and $\bar U(p) = 
U^\dagger(p) \gamma^0$ a $2 \times 4$) such that the equation (\ref{wigvsfd}) can 
be written as:
\be\label{wigvsfd2}
 W(x,k) = \frac{1}{2} \int \frac{\di^3 \p}{\varepsilon} \; \delta^4(k-p) 
 U(p) f(x,p) \bar U(p) - \delta^4(k+p) V(p) \bar f^T (x,p) \bar V(p)
\ee 
The mean value of any operator involving a $4 \times 4$ matrix $A$ such as $\langle
: \Psibar A \Psi: \rangle$ can then be written, using (\ref{wigdir}) and (\ref{wigvsfd2}):
\bea\label{genermat}
&& \langle: \Psibar(x) A \Psi(x): \rangle = \int \di^4 k \; \tr (A W(x,k)) =
 \int \frac{\di^3 \p}{2\varepsilon} \; \left[ \tr (U(p) f(x,p) \bar U(p) A) 
 - \tr (V(p) \bar f^T (x,p) \bar V(p) A) \right] \nonumber \\
&& = \int \frac{\di^3 \p}{2\varepsilon} \; \sum_{rs} f_{rs}(x,p) \bar u(p)_{s} 
  A u(p)_r - \bar f_{rs}(x,p) \bar v_r(p) A v_s(p) \nonumber \\
&&  = \int \frac{\di^3 \p}{2\varepsilon} 
  \; \tr_2 \left( f(x,p) \bar U(p) A U(p) \right) - \tr_2 \left( 
  \bar f^T(x,p) \bar V(p) A V(p) \right) 
\eea 

We have introduced the symbol $\tr_2$ to mean the sum over polarization states (indices
$r,s$), whereas $\tr$ without subscript stands for the trace of $4 \times 4$ spinorial
matrices.

We are now in a position to calculate the mean value of physical densities, like
currents, stress-energy tensor etc. The mean value of the current $\Psibar \gamma^\mu 
\Psi$, is a straightforward application of the eq.~(\ref{genermat}) with $A = \gamma^\mu$:
\begin{eqnarray*}
 j^\mu(x) &\equiv& \langle: \Psibar(x) \gamma^\mu \Psi(x): \rangle = \int 
  \frac{\di^3 \p}{2\varepsilon} \sum_{rs} f_{rs}(x,p) \bar u(p)_{s} 
  \gamma^\mu u(p)_r - \bar f_{rs}(x,p) \bar v_r(p) \gamma^\mu v_s(p) \nonumber \\
 &=& \int \frac{\di^3 \p}{2\varepsilon} \sum_{rs} f_{rs}(x,p) 2 p^\mu \delta_{rs}
  - \bar f_{rs}(x,p) 2 p^\mu \delta_{rs} = 
  \int \frac{\di^3 \p}{\varepsilon} p^\mu \left( \tr_2 f(x,p) - \tr_2 \bar f(x,p)
  \right)
\end{eqnarray*}
where we have used the well known properties of spinors $u,v$. The meaning of the
last expression is clear: the mean value of the current is the momentum integral of
the single-particle distribution in phase space multiplied by their four-velocity,
with a $+$ sign for particles and a $-$ sign for antiparticles. In full accord 
with eq.~(\ref{spectradirac}), $\tr_2 f$ and $\tr_2 \bar f$ are the particle number 
densities in phase space.\\
Similarly, one can obtain the expression of the mean value of the {\em canonical}
stress-energy tensor (see Appendix B for the proof):
\be\label{dcset}
 T^{\mu\nu}(x) \equiv \frac{\ii}{2} \langle: \Psibar(x) \gamma^\mu \codevnu \Psi(x): 
 \rangle = \int \frac{\di^3 \p}{\varepsilon} p^\mu p^\nu \left( \tr_2 f(x,p) +
 \tr_2 \bar f(x,p)\right) 
\ee
and, by using eq.~(\ref{genermat}), the mean value of the {\em canonical} spin 
tensor:
\be\label{dcspint}
 \spt^{\lambda,\mu\nu}(x) \equiv \frac{1}{2} \langle: \Psibar(x) \{ \gamma^\lambda, 
 \Sigma^{\mu \nu} \} \Psi(x): \rangle = \frac{1}{2} \int \frac{\di^3 \p}{2\varepsilon} 
 \tr_2 \left( f(x,p) \bar U(p) \{ \gamma^\lambda, \Sigma^{\mu \nu} \} U(p) \right) 
 - \tr_2 \left( \bar f^T(x,p) \bar V(p) \{ \gamma^\lambda, \Sigma^{\mu \nu}\} 
 V(p) \right)
\ee
where $\Sigma^{\mu\nu} = (\ii/4) [\gamma^\mu,\gamma^\nu]$ are the generators of Lorentz
transformation of spinors. Note that the mean value of the canonical stress-energy tensor 
turns out to be symmetric even though the operator is not. This implies, for the 
free Dirac field, that the divergence of the mean value of the spin tensor (which
equals the antisymmetric part of the mean value of the stress-energy tensor because
of angular momentum conservation) vanishes, that is $\partial_\lambda \langle : 
\wspt^{\lambda,\mu\nu} : \rangle = 0$. This result should have been expected in a 
theory without interaction. We note that this equation holds only if the density 
operator is time-independent, which, as we will see, is not the case for the local 
equilibrium density operator. 

\section{Global thermodynamical equilibrium with rotation}
\label{global}

At global thermodynamical equilibrium with finite angular momentum density the 
density operator is well known \cite{landau,vilenkin} and reads:
\be\label{rhorot}
\wrho = \frac{1}{Z} \exp[ -\widehat H/T + \mu \widehat Q/T + \omegav \cdot 
\widehat {\bf J}/T] \Pro_V
\ee
where $\omegav$ is a constant fixed vector whose physical meaning is that of an 
angular velocity and $T$ is the global temperature, that is the temperature of
a thermostat in contact with the system or that measured by a thermometer at rest
with respect to the external inertial observer. The $\Pro_V$ operator is the 
projector operator onto localized states \cite{becarot2} which is needed in order
to avoid the relativistic singularity at $r = c/\omega$. For this distribution, 
it has been shown the single-particle distribution function of an ideal relativistic 
Boltzmann gas of particles with spin $S$ can be obtained from just statistical 
mechanics arguments \cite{becarot2}:  
\begin{equation}\label{phsp}
  f(x,p)_{rs} = \e^{\xi} \, \e^{-\beta \cdot p}\;\;   
  \frac{1}{2} \left( D^S([p]^{-1} {\sf R}_{\hat \omegav}(\ii \omega/T)[p])+
  D^S([p]^{\dagger} {\sf R}_{\hat \omegav}(\ii \omega/T) [p]^{\dagger-1}) 
  \right)_{rs}
\end{equation}
where $\beta = \frac{1}{T}(1,\omegav \times {\bf x})$ is the inverse temperature 
four-vector; $\lambda$ is the fugacity; $[p]$ is the SL(2,C) matrix corresponding to the Lorentz 
transformation taking the time unit vector $\hat t$ into $\hat p$ (so-called standard
transformation); $D^S$ stands for the $(S,0)$ irreducible representation of SL(2,C) 
(the $(0,S)$ being $D^{S\dagger -1}$ \cite{moussa}); ${\sf R}$ is the SL(2,C) corresponding 
of a rotation, which is calculated for an imaginary angle $\ii \omega/T$. 
Note that, according to the eq.~(\ref{phsp}) the matrix $f$ is not diagonal and:
$$
 \tr_{2S+1} f = \e^{\xi} \, \e^{-\beta \cdot p} \tr_{2S+1} {\sf R}_{\hat \omegav}
 (\ii \omega/T) = \e^{\xi} \, \e^{-\beta \cdot p} \sum_{\sigma=-S}^S \e^{-\sigma \omega/T}
 \equiv \e^{\xi} \, \e^{-\beta \cdot p} \chiomt
$$

We now focus our attention on the $S=1/2$ case. Since Weyl's representation the 
Dirac spinors, with the normalization $\bar u(p)_r u(p)_s = 2m \delta_{rs}$) and 
with $C=\ii \sigma_2$ ($\sigmav$ being Pauli matrices) \cite{moussa,bf2} read:
\be\label{uvspin}
 U(p) = \sqrt{m} {D^S([p]) \choose D^S([p]^{\dagger -1})} \qquad 
 V(p) = \sqrt{m} {D^S([p]C^{-1}) \choose D^S([p]^{\dagger -1}C)}
\ee
and $D^{1/2}({\sf R}_{\hat \omegav}(\ii\omega/T)) = \exp[(-\omega/T) \sigma_3/2]$
one can rewrite the (\ref{phsp}), for spin 1/2 particles in the Boltzmann limit as:
\begin{equation}\label{phspd}
  f(x,p) = \e^\xi \, \e^{-\beta \cdot p}\;\; \frac{1}{2m} \bar U(p) 
  \exp[(\omega/T) \, \Sigma_z] U(p) 
\end{equation}
being:
\be\label{Sigma}
  \Sigma_z = \frac{1}{2} \left( \begin{array}{cc} \sigma_3 \; & \; 0 \\ 
  0 \; & \; \sigma_3 \end{array} \right)
\ee
whereas for antiparticles it can be shown that (see Appendix C):
\begin{equation}\label{phspd2}
  \bar f(x,p) = - \e^{-\xi} \, \e^{-\beta \cdot p}\;\; \frac{1}{2m} \left[ 
  \bar V(p) \exp[-(\omega/T) \, \Sigma_z] V(p) \right]^T 
\end{equation}
The eqs.~(\ref{phspd}) and (\ref{phspd2}) can be written in a fully covariant form 
by introducing the tensor:
\be\label{omegaeq}
  \varpi_{\mu\nu} = (\omega/T) (\delta^1_\mu \delta^2_\nu - \delta^1_\nu \delta^2_\mu)
  = \sqrt{\beta^2} \Omega_{\mu \nu}
\ee
where $\Omega_{\mu\nu}$ turns out to be the acceleration tensor of the Frenet-Serret 
tetrad of the $\beta$ field lines \cite{becarot2}. The above second equality holds 
for a rigid velocity field only \cite{becarot2,becacovstat}. Hence, in the Boltzmann 
limit:
\begin{equation}\label{covphspd}
  f(x,p) = \e^{\xi} \, \e^{-\beta \cdot p}\;\; \frac{1}{2m} \bar U(p) 
  \exp \left[ \frac{1}{2}\varpi^{\mu\nu} \Sigma_{\mu\nu} \right] U(p) \qquad \qquad
 \bar f(x,p) = - \e^{-\xi}\, \e^{-\beta \cdot p}\;\; \frac{1}{2m} \left[ \bar V(p) 
  \exp\left[ -\frac{1}{2}\varpi^{\mu\nu} \Sigma_{\mu\nu}\right] V(p) \right]^T   
\end{equation}

The extension to Fermi-Dirac statistics of this formula should, in principle, be
determined calculating the covariant Wigner function (\ref{wigdir}) with the 
operator (\ref{rhorot}). However, this is not a straightforward calculation like
in the non-rotating case, hence we make an {\em ansatz} about this extension which
reproduces the usual Fermi-Dirac distribution in the non-rotating case and, at the 
same time, has the (\ref{covphspd}) as its Boltzmann limit. We first introduce the 
$4\times 4$ matrices $X,\bar X$ whose relation with $f,\bar f$ reads:
\be\label{xdef}
 \frac{1}{2m} \bar U(p) X(x,p) U(p) = f(x,p) \qquad \qquad
 -\frac{1}{2m} \left[ \bar V(p) \bar X(x,p) V(p) \right]^T = \bar f(x,p)
\ee
and assume that they are, at thermodynamical equilibrium with rotation:
\be\label{mat}
 X = \left( \exp[\beta \cdot p - \xi] \exp\left[ -\frac{1}{2}\varpi : 
 \Sigma\right] + I \right)^{-1} \qquad   
 \bar X = \left( \exp[\beta \cdot p + \xi] \exp\left[ \frac{1}{2}\varpi : 
 \Sigma\right] + I \right)^{-1}
\ee
where $:$ is a shorthand for the rank 2 tensor contraction, i.e. $A:B = A_{\mu\nu}
B^{\mu\nu}$. Consequently, the single-particle distribution functions read:
\bea\label{fdirac}
  f(x,p) &=& \frac{1}{2m} \bar U(p) \left( \exp[\beta \cdot p - \xi] 
    \exp\left[ -\frac{1}{2}\varpi : \Sigma \right] + I \right)^{-1} U(p)  \nonumber \\
 \bar f(x,p) &=& -\frac{1}{2m} \left[ \bar V(p) \left( \exp[\beta \cdot p +\xi] 
    \exp\left[ \frac{1}{2}\varpi : \Sigma \right] + I \right)^{-1} V(p)\right]^T
\eea
which have (\ref{covphspd}) as Boltzmann limit. For the non-rotating case, 
$\varpi=0$, it is easy to check that
$$
 \tr_2 f = 2 \frac{1}{\exp[\beta \cdot p - \xi] + 1}
$$
where the factor $2$ on the right hand side is just the spin degeneracy factor.
The conjectured extension (\ref{mat}) has therefore all the required features
to be a good full quantum statistical formula and, although we cannot provide
a formal proof, it naturally appears as the most natural and almost obvious 
extension.

\section{Local equilibrium of a relativistic fluid of particles with spin}
\label{local}

In quantum relativistic statistical mechanics, local thermodynamical equilibrium
density operator reads:
\be\label{localth}
\wrho_{\rm LE} = \frac{1}{Z_{\rm LE}} \exp\left[ -\int \di^3 \x \; \beta_\nu(x) \wT^{0\nu}(x) 
 - \frac{1}{2} \varpi_{\mu\nu}(x) \wspt^{0,\mu\nu}(x) - \xi(x) \widehat j^0(x) \right]
\ee
and it is obtained by maximizing entropy $S=-\tr [\wrho \log \wrho]$ with the 
constraints of given local values of mean energy-momentum, angular momentum and 
charge density \cite{zubarev}. As entropy is not conserved in nonequilibrium situation, 
the above operator breaks covariance (there cannot be invariant spatial integrals of 
non-conserved currents) and it is time dependent.
This operator is used in derivations of the relativistic Kubo formulae of transport 
coefficients \cite{hosoya} and, in comparison with usual formulations, it has an additional 
term involving the spin tensor, obtained in ref.~\cite{bt2}. In principle, all 
quantities at local thermodynamical equilibrium in quantum relativistic statistical
mechanics should be calculated using (\ref{localth}) as density operator, including
the covariant Wigner function of the Dirac field. However, the full calculation is quite 
complicated and goes beyond the scope of this work. At the lowest order of approximation,
however, we know that it must yield the same formal expression at global thermodynamical
equilibrium with space-time dependent intensive thermodynamics functions, that is 
space-time dependent $\beta$, $\varpi$ and $\xi$. Hence, the single-particle distribution 
functions (\ref{fdirac}) become:
\bea\label{mainform}
  f(x,p) &=& \frac{1}{2m} \bar U(p) \left( \exp[\beta(x)\cdot p - \xi(x)] 
    \exp[-\frac{1}{2}\varpi(x) : \Sigma] + I \right)^{-1} U(p)  \nonumber \\
 \bar f(x,p) &=& -\frac{1}{2m} (\bar V(p) \left( \exp[\beta(x)\cdot p +\xi(x)] 
    \exp[\frac{1}{2}\varpi(x) : \Sigma] + I \right)^{-1} V(p))^T
\eea
The matrices encompassed by the spinors $U, V$ are the corresponding of the $X, \bar X$ 
in eq.~(\ref{mat}) for space-time dependent values of the thermodynamical parameters
and they can be expanded into a power series, each term having an even number of 
$\gamma$ matrices because $\Sigma_{\mu\nu} \propto [\gamma_\mu, \gamma_\nu]$. 
Therefore, the particle phase space densities read:
\bea\label{tr2f}
 \tr_2 f &=& \frac{1}{2m} \tr_2 (\bar U(p) X U(p)) = \frac{1}{2m} 
 \tr (X U(p) \bar U(p)) = \frac{1}{2m} \tr (X (\slashed p + m)) = 
 \frac{1}{2} \tr X \nonumber \\
 \tr_2 \bar f &=& - \frac{1}{2m} \tr_2 (\bar V(p) \bar X V(p)) = 
 - \frac{1}{2m} \tr (\bar X V(p) \bar V(p)) = 
 - \frac{1}{2m} \tr (\bar X (\slashed p - m)) = \frac{1}{2} \tr \bar X 
\eea
where we have used trace cyclicity and the spinorial completeness relations
(written in compact notation):
\be\label{complet}
  U(p) \bar U(p) = (\slashed p + m) I \qquad \qquad 
  V(p) \bar V(p) = (\slashed p - m) I
\ee
Thus, we obtain the free Dirac part of the canonical stress-energy tensor 
(\ref{dcset}) as:
\be\label{dcset2}
  T^{\mu\nu}(x) =\int \frac{\di^3 \p}{\varepsilon} \; p^\mu p^\nu 
  (\tr_2 f + \tr_2 \bar f) = 
  \frac{1}{2} \int \frac{\di^3 \p}{\varepsilon} \; p^\mu p^\nu  
  \left[ \tr X + \tr \bar X \right]
\ee  
and the current:
\be\label{dcch}
 j^\mu(x) = \frac{1}{2} \int \frac{\di^3 \p}{\varepsilon} \; p^\mu 
 \left[ \tr X - \tr \bar X \right]
\ee  
both involving the traces of $X,\bar X$. If $\varpi$, which is adimensional in natural
units, is small enough such that one can write the expansion:
\be\label{xmatexpa}
  \left( \e^{\beta(x)\cdot p \mp \xi(x)} \exp\left[ \mp \frac{1}{2}\varpi(x):\Sigma
  \right] + I \right)^{-1}
  = \sum_{n=1}^\infty (-1)^{n+1} \e^{-n \beta(x)\cdot p \pm n \xi(x)}
  \exp\left[ \pm \frac{n}{2}\varpi(x):\Sigma \right] 
\ee 
then it is possible to obtain an approximate expression of those traces for 
$\varpi_{\mu\nu} \ll 1$. Note that at full rotational equilibrium this condition 
amounts to require $\hbar \omega/KT \ll 1$ (natural units purposely restored) which 
is a normally fulfilled condition. Hence:
\be\label{tracc1}
 \tr \left( \exp\left[ \pm \frac{n}{2}\varpi(x):\Sigma \right] \right) \simeq
  \tr \left( I  \pm \frac{n}{2}\varpi(x):\Sigma + \frac{n^2}{4} 
  \varpi(x):\Sigma \, \varpi(x):\Sigma \right) =
  4 + \frac{n^2}{4} \varpi(x)^{\lambda\rho} \varpi(x)^{\sigma\tau}
  \tr \left( \Sigma_{\lambda\rho} \Sigma_{\sigma\tau} \right)
\ee
where the tracelessness of $\Sigma$ matrices has been used. By using known formulae for
the traces of $\gamma$ matrices it can be shown that:
$$
 \tr \left( \Sigma_{\lambda\rho} \Sigma_{\sigma\tau} \right) =
 g_{\lambda\sigma} g_{\rho\tau} - g_{\lambda\tau} g_{\rho\sigma}
$$
whence the eq.~(\ref{tracc1}) becomes:
\be\label{tracc2}
 \tr \left( \exp\left[ \pm \frac{n}{2}\varpi(x):\Sigma \right] \right) \simeq
  4 + \frac{n^2}{2} \varpi(x):\varpi(x)
\ee
Therefore, we have:
\be\label{traccx}
 \tr X \simeq \sum_{n=1}^\infty (-1)^{n+1} \e^{-n \beta(x)\cdot p + n \xi(x)}
 \left( 4 + \frac{n^2}{2} \varpi(x):\varpi(x) \right) = 
 4 n_F + \frac{1}{2} n_F(1-n_F)(1-2n_F) \varpi(x):\varpi(x)
\ee
where:
$$
 n_F = \frac{1}{\e^{\beta(x)\cdot p - \xi(x)}+1}
$$ 
and similarly for $\bar X$ with the replacement $\xi \to -\xi$. One can then plug
the eq.~(\ref{traccx}) and its corresponding for $\bar X$ into eqs.~(\ref{dcset2})
and (\ref{dcch}) to obtain, e.g., for the charge density:
$$
 j^0(x) = 2 \int \di^3 \p \; (n_F - \bar n_F) + \varpi(x):\varpi(x)
 \frac{1}{4} \int \di^3 \p \; \left[ n_F(1-n_F)(1-2n_F)
 - \bar n_F(1- \bar n_F)(1-2 \bar n_F) \right]
$$
For $\varpi=0$ one recovers the usual expression; it is worth noting that the lowest
order correction to charge density and stress-energy tensor is quadratic in $\varpi$,
i.e. in $\hbar \omega/KT$ at equilibrium.

It is now crucial to know the physical meaning of the tensor $\varpi$, whose expression 
is, as yet, known only at equilibrium. While the general physical meaning of the fields 
$\beta$ and $\xi$ can be easily inferred from the equilibrium limit ($\beta$ is the 
local inverse temperature four-vector field and $\xi$ the ratio between the comoving 
chemical potential $\mu_0(x)$ and the local comoving temperature $T_0(x)=1/\sqrt{\beta^2}$), 
$\varpi_{\mu \nu}(x)$'s expression cannot be uniquely obtained from the equilibrium 
distribution. The reason of this ambiguity is that at rotational equilibrium, the 
tensor $\varpi$ is, at the same time (see eq.~(\ref{omegaeq})):
\be\label{omegaeq2}
  \varpi_{\mu \nu} = \sqrt{\beta^2} \Omega_{\mu \nu}
\ee
where $\Omega_{\mu\nu}$ is the acceleration tensor of the Frenet-Serret tetrad of 
the $\beta$ field lines \cite{becarot2} and \cite{degroot,becacovstat}:
\be\label{omega2}
 \varpi_{\mu\nu} = - \frac{1}{2} (\partial_\mu \beta_\nu - \partial_\nu \beta_\mu)
\ee
In a nonequilibrium situation, the right hand sides of eqs.~(\ref{omega2}) and 
(\ref{omegaeq2}) differ and it is not obvious which one applies, perhaps 
neither. However, if the system is not too far from equilibrium, $\varpi$ cannot be 
too distant from the right hand side of (\ref{omega2}). Particularly, the difference 
must be of the 2nd order in the gradients of the $\beta$ field, i.e.:
\be\label{omega3}
 \varpi_{\mu\nu} = - \frac{1}{2} (\partial_\mu \beta_\nu - \partial_\nu \beta_\mu)
 + {\cal O}(\partial^2 \beta)
\ee
where ${\cal O}(\partial^2 \beta)$ stands for terms which are quadratic in the $\beta$
gradients (for instance: $(\partial \cdot \beta)(\partial_\mu \beta_\nu - \partial_\nu 
\beta_\mu)$) or second order gradients (like $\partial_\mu\partial_\nu \beta^\lambda$)
which vanish at equilibrium, where \cite{degroot,becacovstat}:
$$
 \beta_\mu = b_\mu + \varpi_{\mu \nu} x^\nu
$$
being $b$ and $\varpi$ constants. For the lowest-order formulation of relativistic
hydrodynamics, the expression (\ref{omega3}) is sufficient to determine the expression
of spin-related quantities, as we will see in the next two sections.

\section{The spin tensor}
\label{spintensor}

In order to estimate the polarization of particles, what will be done in the next
section, it is necessary to calculate the spin tensor first. We start by working out
eq.~(\ref{dcspint}) with the help of eq.~(\ref{mainform}) : 
\bea\label{dcspint2}  
 \spt^{\lambda,\mu\nu} &=& \frac{1}{2} \int \frac{\di^3 \p}{2\varepsilon} 
 \; \tr_2 \left( f(x,p) \bar U(p) \{ \gamma^\lambda, \Sigma^{\mu \nu} \} U(p) \right) - 
 \tr_2 \left( \bar f^T(x,p) \bar V(p) \{ \gamma^\lambda, \Sigma^{\mu \nu}\} V(p) \right) \nonumber \\
 &=& \frac{1}{4m} \int \frac{\di^3 \p}{2\varepsilon}  \; \tr_2 \left(\bar U(p) X U(p) \bar U(p) 
 \{ \gamma^\lambda, \Sigma^{\mu \nu} \} U(p) \right) + \tr_2 \left( \bar V(p) \bar X V(p)
 \bar V(p) \{ \gamma^\lambda, \Sigma^{\mu \nu}\} V(p) \right) \nonumber \\
 &=& \frac{1}{4m} \int \frac{\di^3 \p}{2\varepsilon} \; \tr_2 \left(\bar U(p) X (\slashed p + m) 
 \{ \gamma^\lambda, \Sigma^{\mu \nu} \} U(p) \right) + \tr_2 \left( \bar V(p) (\slashed p - m) 
 \bar X \{ \gamma^\lambda, \Sigma^{\mu \nu}\} V(p) \right) \nonumber \\
 &=& \frac{1}{4m} \int \frac{\di^3 \p}{2\varepsilon} \; \tr \left(X (\slashed p + m) 
 \{ \gamma^\lambda, \Sigma^{\mu \nu} \} (\slashed p + m) \right) + \tr \left( \bar X (\slashed p - m) 
 \{ \gamma^\lambda, \Sigma^{\mu \nu}\} (\slashed p - m) \right) \nonumber \\
 &=& \frac{1}{4} \int \frac{\di^3 \p}{2\varepsilon} \; \tr \left(\{ X,\slashed p \} 
 \{ \gamma^\lambda, \Sigma^{\mu \nu} \} \right) - \tr \left(\{ \bar X, \slashed p \} 
 \{ \gamma^\lambda, \Sigma^{\mu \nu} \} \right) 
\eea 
where we have used the relations (\ref{complet}) and taken into account that $X,\bar X$ 
are a linear combination of terms with an even number of $\gamma$ matrices.
By taking advantage of the cyclicity of the trace, the last expression can be rewritten:
$$
\spt^{\lambda,\mu\nu} = \frac{1}{4} \int \frac{\di^3 \p}{2\varepsilon} \; 
 \tr \left(\{ \slashed p, \gamma^\lambda \} \{X, \Sigma^{\mu \nu} \} \right) + 
 \tr \left( [\Sigma^{\mu \nu},\slashed p] [X,\gamma^\lambda] \right) -
 \tr \left(\{ \slashed p, \gamma^\lambda \} \{\bar X, \Sigma^{\mu \nu} \} \right) - 
 \tr \left( [\Sigma^{\mu \nu},\slashed p] [\bar X,\gamma^\lambda] \right)
$$
and, using the known relations:
$$
 \{ \slashed p, \gamma^\lambda \} = 2 p^\lambda \qquad \qquad 
 [\Sigma^{\mu \nu},\slashed p] = - \ii p^\mu \gamma^\nu + \ii p^\nu \gamma^\mu
$$
as:
\begin{eqnarray*}
\spt^{\lambda,\mu\nu} &=& \frac{1}{4} \int \frac{\di^3 \p}{2\varepsilon} \; 
 2 p^\lambda \tr \left(\{X, \Sigma^{\mu \nu} \} \right) - 
 \tr \left( \ii (p^\mu \gamma^\nu - p^\nu \gamma^\mu) [X,\gamma^\lambda] \right) -
 2 p^\lambda \tr \left(\{\bar X, \Sigma^{\mu \nu} \} \right) + 
 \tr \left( \ii (p^\mu \gamma^\nu - p^\nu \gamma^\mu) [\bar X,\gamma^\lambda] 
 \right)
 \nonumber \\
 &=& \frac{1}{4} \int \frac{\di^3 \p}{2\varepsilon} \; 4 p^\lambda 
 \tr \left(X \Sigma^{\mu \nu} \right) - p^\mu \tr \left( \ii [\gamma^\lambda,\gamma^\nu] 
 X \right) + p^\nu \tr \left( \ii [\gamma^\lambda,\gamma^\mu] X \right) \nonumber \\
 && - 4 p^\lambda \tr \left(\bar X \Sigma^{\mu \nu} \right) +  p^\mu 
 \tr \left( \ii [\gamma^\lambda,\gamma^\nu] \bar X \right) - p^\nu 
 \tr \left( \ii [\gamma^\lambda,\gamma^\mu] \bar X \right) 
 \nonumber \\
 &=& \frac{1}{4} \int \frac{\di^3 \p}{2\varepsilon} \; 4 p^\lambda 
 \tr \left(X \Sigma^{\mu \nu} \right) - 4 p^\mu \tr \left( \Sigma^{\lambda\nu} X \right) 
 + 4 p^\nu \tr \left(\Sigma^{\lambda\mu} X \right) - 4 p^\lambda 
 \tr \left(\bar X \Sigma^{\mu\nu} \right) +  4 p^\mu \tr \left( \Sigma^{\lambda\nu} 
 \bar X \right) - 4 p^\nu \tr \left( \Sigma^{\lambda\mu} \bar X \right) 
\end{eqnarray*} 
Finally, we can write the spin tensor as:
\be\label{dcspint3}
 \spt^{\lambda,\mu\nu} = \frac{1}{2} \int \frac{\di^3 \p}{\varepsilon} \; \left( 
 p^\lambda \Theta^{\mu \nu} + p^\nu \Theta^{\lambda\mu} + p^\mu \Theta^{\nu\lambda} + 
 p^\lambda \bar\Theta^{\mu \nu} + p^\nu \bar\Theta^{\lambda \mu} + p^\mu 
 \bar\Theta^{\nu\lambda} \right)
\ee
where
$$
  \Theta^{\mu \nu} \equiv \tr \left(X \Sigma^{\mu \nu} \right)
\qquad \bar\Theta^{\mu \nu} \equiv - \tr \left(\bar X \Sigma^{\mu \nu} 
 \right)
$$
Altogether, the form (\ref{dcspint3}) of the canonical spin tensor only depends 
on the fact that $X,\bar X$ are a superposition of an even number of $\gamma$ matrices. 
The full antisymmetry of the indices is now apparent, although it was already contained 
in the operator definition ensuing from the properties of $\gamma$ matrices. Thanks 
to trace cyclicity, $\Theta$ can be written as a derivative with respect to the 
$\varpi$ tensor:
\bea\label{theta}
 \Theta^{\mu \nu} = \tr ( X \Sigma^{\mu \nu} ) &=& 
 \sum_{n=1}^\infty (-1)^{n+1} \e^{-n \beta(x)\cdot p + n \xi(x)}
 \tr \left( \exp\left[ \frac{n}{2}\varpi(x):\Sigma \right] 
 \Sigma^{\mu \nu} \right) \nonumber \\
 &=& \sum_{n=1}^\infty (-1)^{n+1} \e^{-n \beta(x)\cdot p + n \xi(x)} \frac{1}{n} 
 \frac{\partial}{\partial \varpi_{\mu\nu}} \tr \left( \exp\left[ 
 \frac{n}{2}\varpi(x):\Sigma\right] \right) \nonumber \\
 &=& \frac{\partial}{\partial \varpi_{\mu\nu}} 
 \tr \left( \log \left\{ I + \e^{-\beta(x)\cdot p + \xi(x)} 
 \exp \left[ \frac{1}{2}\varpi(x):\Sigma \right] \right\} \right)
\eea
where the (\ref{xmatexpa}) has been used. Then, using the approximation (\ref{tracc2}),
the eq.~(\ref{theta}) becomes:
\bea\label{theta2}
 \Theta^{\mu \nu} &\simeq& \sum_{n=1}^\infty (-1)^{n+1} \e^{-n \beta(x)\cdot p 
 + n \xi(x)} \frac{1}{n} \frac{\partial}{\partial \varpi_{\mu\nu}} 
 \left( 4 + \frac{n^2}{2} \varpi(x):\varpi(x) \right) \nonumber \\
 &=& \sum_{n=1}^\infty (-1)^{n+1} \e^{-n \beta(x)\cdot p 
 + n \xi(x)} n \varpi(x)^{\mu\nu} = n_F (1 - n_F) \varpi(x)^{\mu\nu}
\eea
Likewise, it can be shown that:
\be\label{thetabar}
  \bar \Theta^{\mu \nu} \simeq \bar n_F (1 - \bar n_F) \varpi(x)^{\mu\nu}
\ee
Note that, according to (\ref{dcspint3}), (\ref{theta2}) and (\ref{thetabar}) the 
spin tensor, at the lowest order, depends linearly on $\varpi$, unlike the current 
and the stress-energy tensor. Now, being:
$$
 \frac{1}{2} \int \frac{\di^3 \p}{\varepsilon} \; p^\lambda \Theta^{\mu \nu}
 = \varpi(x)^{\mu\nu} \frac{1}{2} \int \frac{\di^3 \p}{\varepsilon} \; 
 p^\lambda n_F(p) (1 - n_F(p)) =  - \varpi(x)^{\mu\nu} 
 \frac{1}{2}\frac{\partial}{\partial \beta_\lambda} \int \frac{\di^3 \p}{\varepsilon}
 \; n_F(p)  
$$
and defining the scalar function:
$$
 F(\beta^2,\xi) \equiv \int \frac{\di^3 \p}{\varepsilon} \; n_F(p) =
  \int \frac{\di^3 \p}{\varepsilon} \; \frac{1}{\e^{\beta \cdot p - \xi} +1}
$$
one can write:
\be\label{theta3}
 \frac{1}{2} \int \frac{\di^3 \p}{\varepsilon} \; p^\lambda \Theta^{\mu \nu}
 = - \frac{\partial F}{\partial \beta^2} \beta^\lambda \varpi(x)^{\mu\nu}
\ee
and similarly for the antiparticle term. Therefore, plugging eq.~(\ref{theta3}) 
into (\ref{dcspint3}):
\be\label{dcspint4}
 \spt^{\lambda,\mu\nu} = \iota \, u^\lambda \varpi(x)^{\mu\nu} + {\rm rotation\;
 of\; indices} + \bar\iota \, u^\lambda \varpi(x)^{\mu\nu} + {\rm rotation\;
 of\; indices} 
\ee
where
$$
 \iota \equiv - \frac{1}{T_0} \frac{\partial F}{\partial \beta^2}
$$
and similarly for $\bar \iota$.
The formula (\ref{dcspint4}) extends a formula obtained in ref.~\cite{becarot2}
for the ideal relativistic Boltzmann gas at equilibrium. In fact, in that paper, it
was {\em assumed} that the spin tensor $\spt^{\lambda,\mu\nu}$ could be factorized into 
the product of the four-velocity $u^\lambda$ and an antisymmetric rank 2 tensor 
$\sigma^{\mu\nu}$. We can now see that such an assumption is in general not valid. 
Even in the simplest case of an ideal Fermi gas at equilibrium the spin tensor does 
not factorize. However, the time component of the spin tensor $\wspt^{0,\mu\nu}$ at equilibrium, 
what can be defined as "spin density", can be shown to agree with the expression
found in ref.~\cite{becarot2}. Because of the axial symmetry of the operator (\ref{rhorot}), 
the only non-vanishing component of a rank 2 antisymmetric tensor, such as 
$\Theta$ and $\bar\Theta$ is the $(12)$ or $(21)$, and so from (\ref{dcspint3}): 
\be\label{spinden3}
 \spt^{0,12}(x) = \frac{1}{2} \int \di^3 \p \;  \tr (X \Sigma^{12})
  - \tr (\bar X \Sigma^{12})
\ee
Taking the Boltzmann limits of $X,\bar X$ and rewriting $\Sigma^{12}=\Sigma_z$:
\bea\label{spinden4}
  \spt^{0,12}(x)  &=& \frac{1}{2} \int \di^3 \p \; \e^{-\beta \cdot p}  
  \; \left\{ \e^\xi \, \tr (\exp[(\omega/T) \, \Sigma_z] \Sigma_z)
  - \e^{-\xi} \tr (\exp[- (\omega/T) \, \Sigma_z] \Sigma_z) \right\} \nonumber \\
  &=& \frac{1}{2} \int \di^3 \p \; \e^{-\beta \cdot p}  \; \left\{ \e^\xi \, 
  \frac{\partial}{\partial \omega/T} \tr (\exp[(\omega/T) \, \Sigma_z])
  + \e^{-\xi} \frac{\partial}{\partial \omega/T} \tr (\exp[-(\omega/T) \, \Sigma_z])\right\}
  \nonumber \\
  &=& \frac{1}{2}(\e^\xi + \e^{-\xi}) \int \di^3 \p \; \e^{-\beta \cdot p} \, 
  \frac{\partial}{\partial \omega/T} \tr (\exp[(\omega/T) \, \Sigma_z]) =
  \frac{1}{2} (\e^\xi + \e^{-\xi}) \int \di^3 \p \; \e^{-\beta \cdot p} \, 
  \frac{\partial}{\partial \omega/T} 2 \, \tr (\exp[(\omega/T)\sigma_3/2]) \nonumber \\
  &=& (\e^\xi + \e^{-\xi}) \int \di^3 \p \; \e^{-\beta \cdot p} \, 
  \frac{\partial}{\partial \omega/T} \tr (\exp[(\omega/T)\sigma_3/2]) 
\eea
where the third equality ensues from the invariance of this trace under a change
of sign of $\omega$. This result indeed coincides with the spin density calculated in
ref.~\cite{becarot2} for an ideal relativistic rotating gas at equilibrium, obtained
without the use of quantum fields.

\section{Polarization}
\label{polarization}

The main consequence of the distribution functions (\ref{mainform}) is that spin $1/2$
particles get polarized at local thermodynamical equilibrium. In this section, we will
calculate the polarization four-vector, which, for a particle with mass $m$ and 
four-momentum $p$ is defined as:
\be\label{pol1}
  \Pi_\mu = -\frac{1}{2} \epsilon_{\mu\rho\sigma\tau} S^{\rho\sigma} \frac{p^\tau}{m}
\ee
where $S^{\rho\sigma}$ is the mean value of the {\em total} angular momentum operator of
the single particle. If we want to know the mean polarization vector of a particle with 
momentum $p$ around the space-time point $x$, we cannot but divide the total angular 
momentum density in phase space ${\cal J}^{0,\rho\tau}(x,p)$ by the density of 
particles in phase space, that is:
\be\label{pol2}
 \langle \Pi_\mu (x,p) \rangle = 
 -\frac{1}{2} \frac{1}{\tr_2 f} \epsilon_{\mu\rho\sigma\tau} 
 \frac{\di {\cal J}^{0,\rho\sigma}(x,p)} {\di^3 p} \frac{p^\tau}{m}
\ee
where ${\cal J}(x)$ is the total angular momentum density:
$$
 {\cal J}^{\lambda,\rho\sigma}(x) = x^\rho T^{\lambda \sigma}(x) - x^\sigma 
 T^{\lambda \tau}(x) + \spt^{\lambda,\rho\sigma}(x)
$$
Replacing the stress energy tensor with its expression in (\ref{dcset}):
$$
 \frac{\di {\cal J}^{0,\rho\sigma}(x)}{\di^3 p} = (x^\rho p^\sigma 
 - x^\sigma p^\rho) \tr_2 f(x,p) + \frac{\di \spt^{0,\rho\sigma}(x)}{\di^3 p}
$$
The Levi-Civita tensor in the above equation makes the orbital part of the angular 
momentum density irrelevant, so we are left with the spin tensor contribution only. 
Thus, taking into account that the particle density in phase space is $\tr_2 f = (1/2) 
\tr X$ (see eq.~(\ref{tr2f})) and using (\ref{dcspint3}) for the spin tensor, for 
particles we have: 
\be\label{meanpol}
 \langle \Pi_\mu (x,p) \rangle = 
 -\frac{1}{4 \tr_2 f} \epsilon_{\mu\rho\sigma\tau} \frac{1}{\varepsilon} \; 
 \left( p^0 \Theta^{\rho \sigma} + p^\sigma \Theta^{0\rho} + p^\rho \Theta^{\sigma 0}
 \right) \frac{p^\tau}{m} = -\frac{1}{2 \tr X} \epsilon_{\mu\rho\sigma\tau} 
 \; \Theta^{\rho \sigma} \frac{p^\tau}{m}
\ee
and similarly for antiparticles with $\bar\Theta$ replacing $\Theta$. For particles, 
at the lowest order in $\varpi$, using eqs.~(\ref{theta2}) and eq.~(\ref{traccx}):
\be\label{meanpol2}
 \langle \Pi_\mu \rangle(x,p) \simeq -\frac{1}{8} \epsilon_{\mu\rho\sigma\tau} 
 \; (1 - n_F) \varpi(x)^{\rho\sigma} \frac{p^\tau}{m} 
\ee
For antiparticles, one gets the same formula, with $\bar n_F$ replacing $n_F$. Here
an important comment is in order: the fact that local thermodynamical equilibrium implies
the same orientation for the polarization vector of particles and antiparticles (unlike
e.g. in the electromagnetic field) is a general outcome and does not depend on the 
introduced approximations. It stems from the fact that the spin tensor, as well as 
the angular momentum, is a charge-conjugation even operator, or, more simply stated,
that thermal and mechanical effects do not "see" the internal charge of the particles.
Finally, not far from equilibrium, using (\ref{omega3}), eq.~(\ref{meanpol2}) becomes:
\be
 \langle \Pi_\mu (x,p)\rangle \simeq \frac{1}{16} \epsilon_{\mu\rho\sigma\tau} 
 \; (1 - n_F) \left( \partial^\rho \beta^\sigma - \partial^\sigma \beta^\rho
 \right) \frac{p^\tau}{m} = \frac{1}{8} \epsilon_{\mu\rho\sigma\tau} 
 \; (1 - n_F) \partial^\rho \beta^\sigma \frac{p^\tau}{m}
\ee
We can see that for a degenerate Fermi gas, the polarization tends to zero because
all levels are filled. For a non-degenerate Fermi gas, the formula coincides with
the one obtained for the ideal relativistic Boltzmann gas \cite{becarot1}, as it
should. 

The polarization vector can be separated in its time and space parts as:
\be
 \Pi = (\Pi^0,\Piv) = \frac{1-n_F}{8 m}( (\nabla \times \betav) \cdot {\bf p},
 \varepsilon (\nabla \times \betav ) - \frac{\partial \betav}{\partial t} \times {\bf p}
 - \nabla \beta^0 \times {\bf p})
\ee
The above formula has the remarkable consequence that quasi-free particles get 
polarized not only in a vorticous flow (what was pointed out in previous works 
\cite{becarot1}), but also in a steady temperature gradient without velocity flow, 
i.e. when $\nabla \beta^0 \ne 0$. Restoring natural constants, in the non-relativistic 
limit where the distinction between comoving and observed temperature can be 
neglected, the predicted polarization reads
$$
 \Pi = (\Pi^0,\Piv) = (1-n_F) \frac{\hbar p}{8 m K T^2} (0,\nabla T \times {\hat {\bf p}})  
$$
which is usually tiny but could be relevant in some extreme situations.

Now, the polarization 3-vector $\Piv_0$ in the rest frame of the particle 
with four-momentum $p$ can be found by Lorentz-boosting the above four-vector 
\cite{jackson}:
\be
 \Piv_0 = \Piv - \frac{{\bf p}}{\varepsilon(\varepsilon + m)} \Piv \cdot {\bf p}
\ee
implying a longitudinal polarization (helicity):
\be
 \Piv_0 \cdot {\bf\hat p} = \Piv \cdot {\bf\hat p} - 
 \frac{\p^2}{\varepsilon(\varepsilon + m)} \Piv \cdot {\bf \hat p} = 
 \left( 1 - \frac{\p^2}{\varepsilon(\varepsilon + m)} \right)
 \frac{1-n_F}{8m} \varepsilon (\nabla \times \betav) \cdot {\bf\hat p} =
 \frac{1-n_F}{8} (\nabla \times \betav ) \cdot {\bf\hat p}
\ee 
Hence, the mean helicity of a spin $1/2$ particle is approximately proportional to 
the curl of the $\betav = \gamma{\bf v}/T_0$ field.  
 
It may be of interest, e.g. for relativistic heavy ion collisions, to calculate 
the space-integrated mean polarization vector. For a three-dimensional spacelike 
hypersurface $\Sigma$, one has, using (\ref{dcspint3}):
\be\label{meanpoli} 
  \langle \Pi_\mu (p) \rangle \equiv \frac{\displaystyle \int \di \Sigma_\lambda \; 
  \frac{p^\lambda}{\varepsilon} (-1/2) \epsilon_{\mu\rho\sigma\tau} 
  \frac{\di \spt^{0,\rho\sigma}}{\di^3 \p} \frac{p^\tau}{m}}{\displaystyle 
  \int \di \Sigma_\lambda \; \frac{p^\lambda}{\varepsilon} \tr_2 f(x,p)}
  = - \frac{1}{4} \epsilon_{\mu\rho\sigma\tau} \frac{p^\tau}{m} 
  \frac{\displaystyle \int \di \Sigma_\lambda \; p^\lambda \; \Theta^{\rho \sigma}}
  {\displaystyle \varepsilon \frac{\di N}{\di^3 \p}} 
\ee
Furthermore, taking into account the particle phase space density is $\tr_2 f =
(1/2) tr X$ and using (\ref{traccx}) at the lowest order in $\varpi$:
$$
  \varepsilon \frac{\di N}{\di^3 \p} = \int \di \Sigma_\lambda \; p^\lambda 
  \tr_2 f \simeq 2 \int \di \Sigma_\lambda \; p^\lambda n_F
$$
one finally obtains, using eqs.~(\ref{theta2}) and (\ref{omega2}):
\be 
  \langle \Pi_\mu (p) \rangle \simeq - \frac{1}{4} \epsilon_{\mu\rho\sigma\tau} 
   \frac{p^\tau}{m} \frac{\displaystyle \int \di \Sigma_\lambda \; p^\lambda \; 
   n_F (1-n_F) \varpi^{\rho \sigma}}{\displaystyle \varepsilon \frac{\di N}{\di^3 \p}}
   \simeq \frac{1}{8} \epsilon_{\mu\rho\sigma\tau} \frac{p^\tau}{m} 
   \frac{\displaystyle \int \di \Sigma_\lambda \; p^\lambda \; 
   n_F (1-n_F) \partial^\rho \beta^\sigma}{\displaystyle \int \di \Sigma_\lambda 
   \; p^\lambda n_F}
\ee

These results deserve some discussion concerned with the physical meaning of the spin 
tensor, which has been the crucial ingredient to obtain (\ref{meanpol}) and the ensuing
formulae. In principle, the definition of the polarization vector (\ref{pol1}) involves
the {\em total} angular momentum of the particle, hence the formula should be invariant 
under a change of the stress-energy tensor and spin tensor operators keeping the 
integral of the total angular momentum density invariant, a so-called {\em pseudo-gauge} 
transformation of the stress-energy tensor \cite{hehl} (see also detailed discussion 
in refs.~\cite{bt1,bt2}). However, in a formula like (\ref{meanpol}), one defines 
the local value of polarization and, therefore, a dependence on the particular spin 
tensor is implied. It could be expected that a space-integrated expression like 
(\ref{meanpoli}) would be independent of the spin tensor choice, in fact this is not
the case because of the explicit time-dependence of the local equilibrium density 
operator (see discussion in sect.~\ref{local}). Such a dependence does not enable to
write the mean value of the divergence of an operator as the divergence of its mean
value, thus breaking the pseudo-gauge invariance of the total angular momentum.
For instance, had we used the Belinfante symmetrized stress-energy tensor, the 
ensuing value of polarization at local thermodynamical equilibrium (\ref{meanpoli})   
would vanish. To summarize, the choice of a specific spin tensor operator is necessary 
to calculate the polarization of particles and we have chosen the canonical spin 
tensor (see eq.~(\ref{dcspint}), which is the same used in ref.~\cite{landau2} to 
calculate the polarization of electrons. Even though it might appear disturbing
that polarization at local thermodynamical equilibrium depends on the particular
quantum spin tensor (whence the stress-energy tensor) of the theory, it has been recently
shown that in thermodynamics this is a general feature \cite{bt1,bt2}.

\section{Summary and conclusions}
\label{conclu}

In summary, we have obtained the form of the single-particle invariant distribution
function for particles and antiparticles with spin $1/2$, generalizing the quantum
statistics versions of the J\"uttner distribution:
\bea\label{summary1}
  f(x,p)_{rs} &=& \frac{1}{2m} \bar u_r(p) \left( \exp[\beta(x)\cdot p - \xi(x)] 
    \exp\left[ -\frac{1}{2}\varpi(x)_{\mu\nu}\Sigma^{\mu\nu} \right] + I \right)^{-1} u_s(p)  
    \nonumber \\
 \bar f(x,p)_{rs} &=& -\frac{1}{2m} \bar v_s(p) \left( \exp[\beta(x)\cdot p +\xi(x)] 
    \exp\left[ \frac{1}{2}\varpi(x)_{\mu\nu}\Sigma^{\mu\nu} \right] + I \right)^{-1} v_r(p)
\eea
where $\beta$ is the inverse temperature four-vector, $\xi$ is the ratio between 
comoving chemical potential and temperature, $\Sigma^{\mu\nu} = (\ii/4) [\gamma^\mu,\gamma^\nu]$ 
are the generators of Lorentz transformations of 4-components spinors, $u(p)$ and
$v(p)$ are the spinors solutions of the free Dirac equation and: 
\be\label{summary2}
 \varpi_{\mu\nu} = - \frac{1}{2} (\partial_\mu \beta_\nu - \partial_\nu \beta_\mu)
 + {\cal O}(\partial^2 \beta)
\ee
This formula leads to a generalization of the Cooper-Frye formula for particles with
spin:
\be\label{summary3} 
 \varepsilon \frac{\di N_{rs}}{\di^3 \p} = \int_\Sigma \di \Sigma_\mu p^\mu f_{rs}(x,p)
\ee

The distribution function (\ref{summary1}) implies that spin $1/2$ particles and 
antiparticles have a polarization:
\be\label{summary4}
 \langle \Pi_\mu (x,p)\rangle \simeq \frac{1}{8} \epsilon_{\mu\rho\sigma\tau} 
 \; (1 - n_F) \partial^\rho \beta^\sigma \frac{p^\tau}{m}
\ee 
where $n_F$ is the usual Fermi-J\"uttner distribution. Therefore, particles acquire
a polarization in vorticous flows, which has a longitudinal component and a transverse
polarization even in steady temperature gradients without flow. As this is essentially
a thermo-mechanical effect, one of its distinctive features is that polarization vector 
has the same orientation for both particles and antiparticles, regardless of the 
internal charge.

\section*{Acknowledgments}

This work was partly supported by HIC for FAIR and the Italian Ministery of Research 
and Education PRIN 2009. We thank L. Csernai for interesting discussions and for
useful suggestions.



\appendix

\section*{APPENDIX A - Distribution function for free particles}

To prove eq.~(\ref{spectradirac}), we first manipulate (\ref{wigvsfd}) and (\ref{wigvsfd.2})
multiplying both sides by $\bar u(k)_c$ on the left and $u(k)_d$ on the right:
\begin{eqnarray}\label{appa1}
&& \bar u(k)_c W^+(x,k) u(k)_d = \frac{1}{2} 
 \int \frac{\di^3 \p}{\varepsilon} \; \delta^4(k-p) \sum_{r,s} \bar u(k)_c 
  u_{r}(p) f_{rs}(x,p) \bar u_{s}(p) u(k)_d = 
  2m^2 \int \frac{\di^3 \p}{\varepsilon} \; \delta^4(k-p) f_{cd}(x,p) \nonumber \\
&& = 4m^2 \int \di^4 p \; \delta(p^2-m^2) \theta(p^0) \; \delta^4(k-p) f_{cd}(x,p)
 = 4m^2  \delta(k^2-m^2) \theta(k^0) f_{cd}(x,k)    
\end{eqnarray}
Acting similarly on (\ref{wigvsfd.2}) we get:
\be\label{appa2}
 \bar v(-k)_c W^-(x,k) v(-k)_d = - 4m^2  \delta(k^2-m^2) \theta(-k^0) \bar f_{cd}(x,-k) 
\ee
We now need the usual expansion of the free Dirac field (with the adopted normalizations,
see sect.~\ref{kinetics}):
$$
 \Psi(x) = \sum_{r=1}^2 \frac{1}{(2 \pi)^{3/2}} \int \frac{\di^3 \p}{2\varepsilon} 
 \; \left( u_r(p) \e^{-\ii p \cdot x} a_r(p) + v_r(p) \e^{\ii p \cdot x} b_r^\dagger(p) 
 \right) 
$$
from which we can calculate the covariant Wigner function according to formula
(\ref{wigdir}):
\begin{eqnarray*}
 && W(x,k)_{AB} = \frac{1}{(2\pi)^4} \int \di^4 y \; \e^{-\ii k \cdot y} \langle : 
   \Psibar_B (x+y/2) \Psi_A (x-y/2): \rangle \\
 && = \frac{1}{(2\pi)^4} \int \di^4 y \; \e^{-\ii k \cdot y} \sum_{r,s=1}^2 
  \frac{1}{(2 \pi)^{3}} \int \frac{\di^3 \p'}{2\varepsilon'}  \langle : 
  \; \left( \bar u_{sB}(p') \e^{\ii p' \cdot (x+y/2) } a^\dagger_s(p') + 
  \bar v_{sB}(p') \e^{-\ii p' \cdot (x+y/2)} b_s^(p') \right) \\ 
  && \times \int \frac{\di^3 \p}{2\varepsilon}  \left( u_{rA}(p) 
  \e^{-\ii p \cdot (x-y/2)} a_r(p) + v_{rA}(p) \e^{\ii p \cdot (x-y/2)} b_{r}^\dagger(p)
  \right) :\rangle 
\end{eqnarray*}
Integrating the $y$ variable, the last expression becomes:
\begin{eqnarray*}
&& W(x,k) = \sum_{r,s=1}^2  \frac{1}{(2 \pi)^{3}} 
  \int \frac{\di^3 \p'}{2\varepsilon'} \int \frac{\di^3 \p}{2\varepsilon}\\
&& \left(  u_r(p) \bar u_s(p') \langle : a^\dagger_s(p') a_r(p) : \rangle  
 \e^{\ii (p'-p) \cdot x} \delta^4( k - p'/2 - p/2) 
 +  v_r(p) \bar u_s(p') \langle : a^\dagger_s(p') b^\dagger_r(p) : \rangle 
 \e^{\ii (p'+ p) \cdot x} \delta^4( k - p'/2 + p/2) \right. \\
 && \left. +  u_r(p)\bar v_s(p') \langle : b_s(p') a_r(p) : \rangle \e^{- \ii (p'+ p) \cdot x}
  \delta^4( k + p'/2 - p/2) + v_r(p) \bar v_s(p') \langle : b_s(p') b^\dagger_r(p) : \rangle  
 \e^{- \ii (p'- p) \cdot x}\delta^4 (k + p'/2 + p/2)  \right)
\end{eqnarray*}
If we now integrate in $\di^3 \x$, we get:
\begin{eqnarray*}
&& \int \di^3 \x \; W(x,k) = \sum_{r,s=1}^2   
  \int \frac{\di^3 \p'}{2\varepsilon'} \int \frac{\di^3 \p}{2\varepsilon}
  \left(  u_r(p) \bar u_s(p) \langle : a^\dagger_s(p) a_r(p) : \rangle 
   \delta^3({\bf p}-{\bf p}') \delta^4(k - p) \right. \\
 && \left. + v_r(p) \bar u_s(p') \langle : a^\dagger_s(p') b^\dagger_r(p) : \rangle 
 \delta^3({\bf p}+{\bf p}') \delta^3({\bf k} + {\bf p}) \delta(k^0) \e^{ 2 \ii p^0 t} 
  +  u_r(p)\bar v_s(p') \langle : b_s(p') a_r(p) : \rangle \delta^3({\bf p}+{\bf p}')
 \delta^3({\bf k} - {\bf p}) \delta(k^0) \e^{ - 2 \ii p^0 t}  \right. \\
 && \left. + v_r(p) \bar v_s(p) \langle : b_s(p) b^\dagger_r(p) : \rangle \delta^3({\bf p}-{\bf p}')
  \delta^4 (k + p)  \right)
\end{eqnarray*}
The two terms with $\delta(k^0)$ are contact terms which can be neglected. Integrating
again in $\p'$:
\begin{eqnarray*}
&& \int \di^3 \x \; W(x,k) = \sum_{r,s=1}^2   
  \int \frac{\di^3 \p}{4\varepsilon^2} \left(  u_r(p) \bar u_s(p) 
  \langle : a^\dagger_s(p) a_r(p) : \rangle \delta^4(k - p)
 + v_r(p) \bar v_s(p) \langle : b_s(p) b^\dagger_r(p) : \rangle \delta^4 (k + p) \right) \\
 && = \sum_{r,s=1}^2 \int \di^4 p \; \frac{1}{2\varepsilon_p} \delta(p^2-m^2) \theta(p^0)
 \left[ u_r(p) \bar u_s(p) \langle : a^\dagger_s(p) a_r(p) : \rangle \delta^4(k - p)
 - v_r(p) \bar v_s(p) \langle : b^\dagger_r(p) b_s(p) : \rangle \delta^4 (k + p) \right] \\
 && =  \delta(k^2-m^2) \frac{1}{2\varepsilon_k}\sum_{r,s=1}^2 \left[ \theta(k^0) u_r(k) \bar u_s(k) 
 \langle : a^\dagger_s(k) a_r(k) : \rangle - \theta(-k^0) v_r(-k) \bar v_s(-k) 
 \langle : b^\dagger_r(-k) b_s(-k) : \rangle \right]
\end{eqnarray*}
Finally, multiplying by $\theta(\pm k^0)$ and the spinors $u$, $v$, we get:
\begin{eqnarray*}\label{appa3}
 &&  \int \di^3 \x \; \bar u(k)_c W^+(x,k) u(k)_d = 4m^2 \theta (k^0) \delta(k^2-m^2)
   \frac{1}{2 \varepsilon_k} \langle : a^\dagger_d(k) a_c(k) : \rangle \\
 && \int \di^3 \x \; \bar v(-k)_c W^-(x,k) v(-k)_d = - 4m^2 \theta (- k^0)
 \delta(k^2-m^2) \frac{1}{2 \varepsilon_k} \langle : b^\dagger_d(-k) b_c(-k) : \rangle 
\end{eqnarray*}
Comparing (\ref{appa3}) with (\ref{appa1}) and (\ref{appa2}), we obtain the 
equation (\ref{spectradirac}).

\section*{APPENDIX B - Canonical stress-energy tensor of the free Dirac field}

We want to prove eq.~(\ref{dcset}). Let us start from:
\be\label{appb1}
 \int \frac{\di^3 \p}{\varepsilon} p^\mu p^\nu \left( \tr_2 f(x,p) +
 \tr_2 \bar f(x,p)\right) = \int \di^4 k \; \tr (\gamma^\mu W(x,k)) k^\nu
\ee
which follows from (\ref{wigvsfd2}) and (\ref{genermat}) with $A=\gamma^\mu$.
The right hand side of the above equation can be also written, by using the covariant 
Wigner function definition (\ref{wigdir}) as:
\be\label{stet}
 \int \di^4 k \; \tr (\gamma^\mu W(x,k)) k^\nu = \frac{\ii}{(2\pi)^4} \sum_{AB} 
 \int \di^4 y \; \frac{\partial}{\partial y_\nu} \e^{-\ii k \cdot y} \langle : 
 \Psibar_B (x+y/2) \Psi_A (x-y/2): \rangle \gamma_{BA}  
\ee
The integral can be worked out by parts:
\begin{eqnarray*}
 && \int \di^4 k \int \di^4 y \; \frac{\partial}{\partial y_\nu} \e^{-\ii k \cdot y} 
 \Psibar_B (x+y/2) \Psi_A (x-y/2) \nonumber \\
 && = \int \di^4 k \int \di \Sigma \; n^\nu \e^{-\ii k \cdot y} 
 \Psibar_B (x+y/2) \Psi_A (x-y/2) - \int \di^4 k \int \di^4 y \; \e^{-\ii k \cdot y} 
 \frac{\partial}{\partial y_\nu} \left( \Psibar_B (x+y/2) \Psi_A (x-y/2) \right) 
 \nonumber \\
 && = (2\pi)^4 \int \di \Sigma \; n^\nu \delta^4(y) \Psibar_B (x+y/2) \Psi_A (x-y/2) - 
 (2\pi)^4 \int \di^4 y \; \delta^4(y) \frac{\partial}{\partial y_\nu} 
 \left( \Psibar_B (x+y/2) \Psi_A (x-y/2) \right) \nonumber \\
 && = 0 - \frac{(2\pi)^4}{2} \left( \partial_\nu \Psibar_B (x) \Psi_A (x) - \Psibar_B (x) 
 \partial_\nu \Psi_A (x) \right)
\end{eqnarray*}
Therefore, by plugging this solution in (\ref{stet}):
$$
 \int \di^4 k \; \tr (\gamma^\mu W(x,k)) k^\nu = \frac{\ii}{2} \langle : \Psibar(x)
 \gamma^\mu \codevnu \Psi(x) : \rangle 
$$
which, together with (\ref{appb1}), proves eq.~(\ref{dcset}).

\section*{APPENDIX C - Distribution function for antiparticles of spin 1/2}

The goal of this Section is to prove that eq.~(\ref{phspd2}) is in fact the Boltzmann 
limit of single-particle distribution function of antifermions with spin 1/2.
The task to be accomplished is to work out the expression:
$$
 -\left[ \bar V(p) \exp[-\omega/T \, \Sigma_z] V(p) \right]^T
$$
Henceforth, we can omit the $D^{1/2}$ symbol, as we work in the fundamental representation
of SL(2,C). According to $V(p)$ definition in eq.~(\ref{uvspin}), and using the definition 
of $\Sigma_z$ (see eq.~(\ref{Sigma}) and the equality ${\sf R}_{\hat \omegav}(\ii\omega/T) 
= \exp (\omega/T \sigma_3/2)$:
\bea\label{C1}
 && -\left[ \bar V(p) \exp[-\omega/T \, \Sigma_z] V(p) \right]^T = 
 - V(p)^T \exp[-\omega/T \, \Sigma_z] \bar V(p)^{T} \nonumber \\
 && = - m (C^{-1 T} [p]^T, C^T [p]^{\dagger -1 T})
 \left( \begin{array}{cc} {\sf R}_{\hat \omegav}(\ii\omega/T)^{-1} \; & \; 0 \\ 
  0 \; & \; {\sf R}_{\hat \omegav}(\ii\omega/T)^{-1} \end{array} \right)
  {([p]^{-1 T} C^{\dagger T} \choose [p]^{\dagger T} C^{\dagger -1 T})} \nonumber \\
 && = - m \left( C^{-1 T} [p]^T {\sf R}_{\hat \omegav}(\ii\omega/T)^{-1}
 [p]^{-1T} C^{\dagger T} + C^T [p]^{\dagger -1 T}{\sf R}_{\hat \omegav}(\ii\omega/T)^{-1}
 [p]^{\dagger T} C^{\dagger -1 T} \right) 
\eea
We can now take advantage of the properties of $C=\ii \sigma_2$, namely:
$$
  C^{T} = C^{\dagger} = C^{-1} = - C \qquad \qquad 
  C A C^{-1} = A^{-1T} \; \; \forall A \in SL(2,C)
$$
so that the eq.~(\ref{C1}) becomes:
\bea\label{C2}
&& - m \left( C^{-1 T} [p]^T {\sf R}_{\hat \omegav}(\ii\omega/T)^{-1}
 [p]^{-1 T} C^{\dagger T} + C^T [p]^{\dagger -1 T}{\sf R}_{\hat \omegav}(\ii\omega/T)^{-1}
 [p]^{\dagger T} C^{\dagger -1 T} \right) \nonumber \\ 
&& = m \left( C [p]^T {\sf R}_{\hat \omegav}(\ii\omega/T)^{-1}
 [p]^{-1 T} C^{-1} + C [p]^{\dagger -1 T}{\sf R}_{\hat \omegav}(\ii\omega/T)^{-1}
 [p]^{\dagger T} C^{-1} \right) \nonumber\\ 
&& = m \left( [p]^{-1} {\sf R}_{\hat \omegav}(\ii\omega/T)^{T}
 [p] + [p]^{\dagger} {\sf R}_{\hat \omegav}(\ii\omega/T)^{T}
 [p]^{\dagger -1} \right) = m \left( [p]^{-1} {\sf R}_{\hat \omegav}(\ii\omega/T)
 [p] + [p]^{\dagger} {\sf R}_{\hat \omegav}(\ii\omega/T)
 [p]^{\dagger -1} \right)
\eea
where, in the last equality, we have used the symmetry of the matrix 
${\sf R}_{\hat \omegav}(\ii \omega/T)$. Therefore, we have shown that:
$$
 -\left[ \bar V(p) \exp[-\omega/T \, \Sigma_z] V(p) \right]^T = 
 m \left( [p]^{-1} {\sf R}_{\hat \omegav}(\ii \omega/T)
 [p] + [p]^{\dagger} {\sf R}_{\hat \omegav}(\ii \omega/T)
 [p]^{\dagger -1} \right)
$$
whence the eq.~(\ref{phspd2}) for antiparticles of spin 1/2 can be also written
as:
\be
  \bar f(x,p)_{rs} = \e^{-\xi} \, \e^{-\beta \cdot p}\;\;   
  \frac{1}{2} \left( [p]^{-1} {\sf R}_{\hat \omegav}(\ii \omega/T)[p])+
  [p]^{\dagger} {\sf R}_{\hat \omegav}(\ii \omega/T) [p]^{\dagger-1}) 
  \right)_{rs}
\ee
This equation is a special case (for $S=1/2$) of the general expression (\ref{phsp}) 
which was obtained in ref.~\cite{becarot2} by means of statistical mechanics arguments.

\end{document}